\documentclass[sigconf]{acmart}
\usepackage{array}
\usepackage{url}
\usepackage{subfig}
\usepackage{multirow,balance}
\usepackage{xcolor}

\copyrightyear{2021}
\acmYear{2021}
\setcopyright{rightsretained}
\acmConference[MM '21]{Proceedings of the 29th ACM International Conference on Multimedia}{October 20--24, 2021}{Virtual Event, China}
\acmBooktitle{Proceedings of the 29th ACM International Conference on Multimedia (MM '21), October 20--24, 2021, Virtual Event, China}\acmDOI{10.1145/3474085.3475587}
\acmISBN{978-1-4503-8651-7/21/10}

\begin{document}
\fancyhead{}

\title{Is Someone Speaking? Exploring Long-term Temporal Features for Audio-visual Active Speaker Detection}

\author{Ruijie Tao, Zexu Pan, Rohan Kumar Das, Xinyuan Qian, Mike Zheng Shou, Haizhou Li}
\email{{ruijie.tao, pan_zexu}@u.nus.edu, ecerohan@gmail.com, 
{eleqian, mikeshou, haizhou.li}@nus.edu.sg}
\affiliation{
  \institution{Department of Electrical and Computer Engineering, National University of Singapore}
  \city{Singapore}
  \country{Singapore}
}

\begin{abstract}

Active speaker detection (ASD) seeks to detect who is speaking in a visual scene of one or more speakers. The successful ASD depends on accurate interpretation of short-term and long-term audio and visual information,  as well as audio-visual interaction. Unlike the prior work where systems make decision instantaneously using short-term features, we propose a novel framework, named TalkNet, that makes decision by taking both short-term and long-term features into consideration. TalkNet consists of audio and visual temporal encoders for feature representation, audio-visual cross-attention mechanism for inter-modality interaction, and a self-attention mechanism to capture long-term speaking evidence. The experiments demonstrate that TalkNet achieves 3.5\% and 2.2\% improvement over the state-of-the-art systems on the AVA-ActiveSpeaker dataset and Columbia ASD dataset, respectively. Code has been made available at: \textcolor{magenta}{\url{https://github.com/TaoRuijie/TalkNet_ASD}}.

\end{abstract}




\maketitle

\section{Introduction}

\begin{figure}[h]
  \centering
  \includegraphics[width=\linewidth]{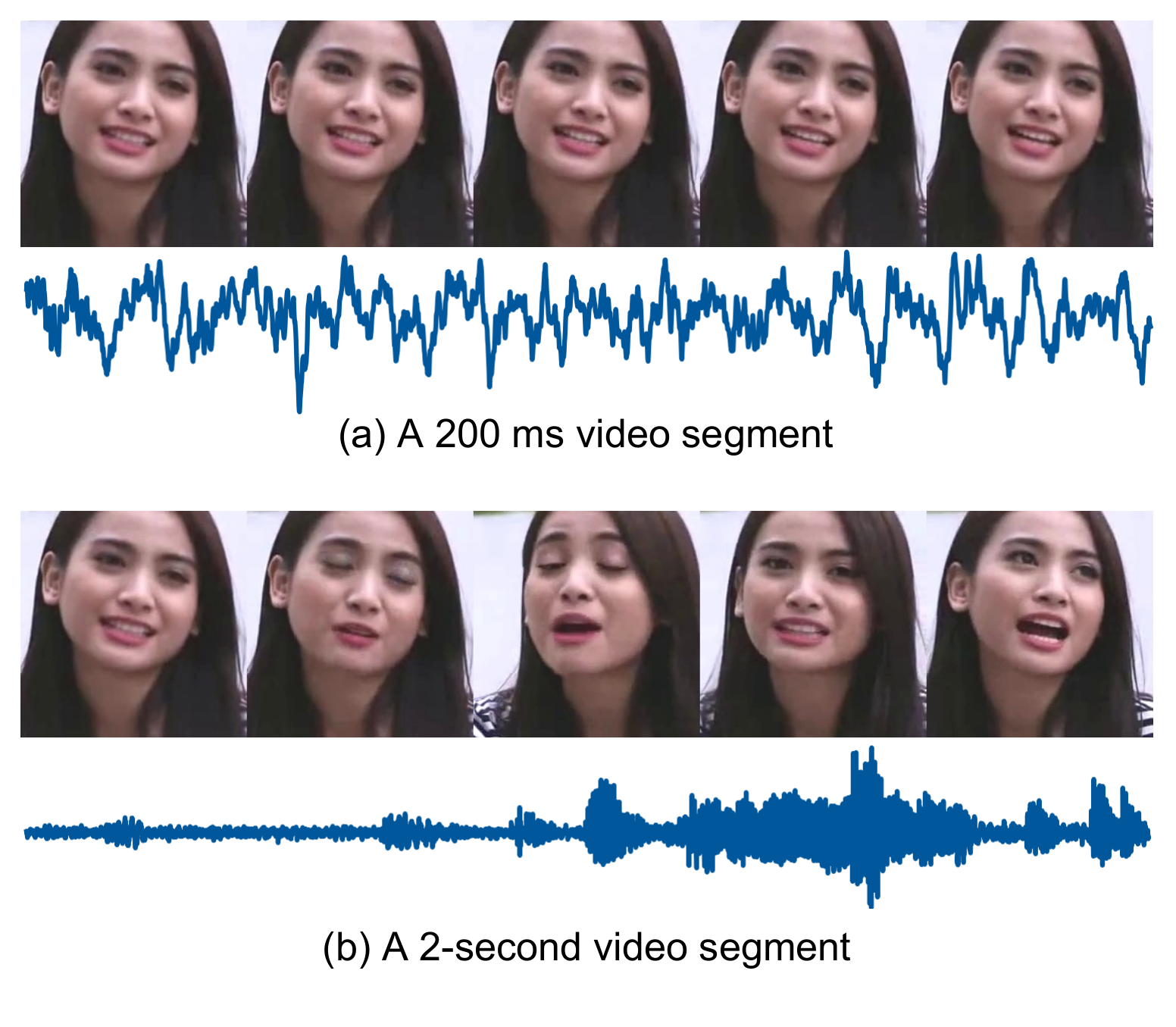}
  \caption{An illustration of the speaking evidence between a short segment and a long segment. (a) Five video frames evenly sampled from a 200 ms video segment, where the speaking activity is not evident. (b) Five video frames evenly sampled from a 2-second video segment, where the speaking activity becomes evident in long-term temporal context and audio-visual synchronization.}
  \label{fig:Active speaker detection}
\end{figure}

Active speaker detection (ASD) seeks to detect who is speaking in a visual scene of one or more speakers~\cite{roth2020ava}. As the speaking circumstances are very fluid and change dynamically, ASD has to predict at a fine granularity in time, i.e., at video frame level. This task is an essential frontend for a wide range of multi-modal applications such as audio-visual speech recognition~\cite{afouras2018deep}, speech separation~\cite{pan2020muse}, speaker diarization~\cite{chung2020spot, chung2019said} and speaker tracking~\cite{qian2021audio, qian2021multi}.

Among the factors, that humans judge whether a person is speaking, are 1) Does the audio of interest belong to human voice? 2) Are the lips of the person of interest moving? 3) If the above are true, is the voice synchronized with the lip movement? Based on this cognitive finding, there have been deep learning solutions that extract audio and visual features to make binary classification~\cite{tao2017bimodal, ariav2019end, chung2016out, roth2020ava}. Despite much progress, the existing ASD systems have not fully benefited from two aspects of available information: the temporal dynamics of audio and visual flow, and the interaction between audio and visual signals, that limits the scope of applications, especially for challenging real-world scenarios. 

As the short-term audio and visual features represent the salient cues for ASD, most of the existing studies are focused on segment-level information, e.g., a video segment of 200 to 600 ms. However, as illustrated in Figure ~\ref{fig:Active speaker detection}(a), it is hard to judge the speaking activity from a video segment of 200 ms, not to mention the audio-visual synchronization. A longer 2-second video, as displayed in Figure~\ref{fig:Active speaker detection}(b),  would be more evident of the speaking episode. When humans are detecting an active speaker, we typically consider saying an entire sentence that spans over hundreds of video frames for a decision, for example, an audio-visual episode lasting 5 seconds contains 15 words on average~\cite{cutts2020oxford, tauroza1990speech}. A short-term segment of 200 ms doesn't even cover a complete word. Furthermore, single modality embedding is not reliable in some challenging scenarios. For example, the voice we hear might come from someone else than the target speaker; at the same time, there could be false lip movements, e.g., laughing, eating, and yawning, that are not related to speaking. To summarize, we consider that the inter-modality synchronization, such as speech-lip, speech-face, over the span of an utterance provides more reliable information than short-term segments. 

The systems with short-term features extract audio-visual embedding from a fixed-length short segment, e.g., 200 ms~\cite{chung2016out, chung2019naver, zhangmulti}, 300 ms~\cite{tao2017bimodal}, and 440 ms segment~\cite{alcazar2020active, leon2021maas}. By simply increasing the segment size, we are getting the average properties of the segment at the cost of the time resolution of speaking activities. A better way to capture the long-term temporal context is to encode the history of audio or video frame sequence. In this paper, we study an audio-visual ASD framework, denoted as TalkNet. For video signals, the minimum unit is a video frame, i.e., a static image. We study a temporal network to encode the temporal context over multiple video frames. For audio signals, the minimum unit is an audio frame of tens of milliseconds. We study an audio temporal encoder to encode the temporal context over multiple audio frames. In terms of backend classifier, we study an audio-visual cross-attention mechanism to capture inter-modality evidence and a self-attention mechanism to capture long-term speaking evidence.  

To the best of our knowledge, this paper is the first study on the use of long-term temporal context, and audio-visual inter-modality interaction for ASD. We make the following contributions.
\begin{itemize}
\item We propose a feature representation network to capture the long-term temporal context from audio and visual cues;
\item We propose a backend classifier network that employs audio-visual cross-attention, and self-attention to learn the audio-visual inter-modality interaction;
\item We propose an effective audio augmentation technique to improve the noise-robustness of the model.
\end{itemize}

The rest of the paper is organized as follows. In Section~\ref{secii}, we discuss the related work. In Section~\ref{seciii}, we formulate the proposed TalkNet framework and present its training process. In Section~\ref{seciv} and Section~\ref{secv}, we report the experiments and their results, respectively. Finally, Section~\ref{conc} concludes the study. 

\section{Related work}
\label{secii}

This research is built on prior studies on the detection of audio and visual events and the modeling long-term temporal dynamics of audio-visual signals.

\subsection{Active Speaker Detection}
% ASD based on single modality
There have been prior studies on ASD using audio, video, and the fusion of both. In voice activity detector (VAD), we study how  to detect the presence of speech as opposed to other acoustic noises~\cite{sehgal2018convolutional, ding2019personal}. However, in real-world scenarios, audio signals by distant microphones are inherently ambiguous because of the overlapping speech and the corruption from background noise, which poses challenges to the VAD task. For vision, the facial~\cite{patrona2016visual} and upper-body~\cite{shahid2021s, chakravarty2015s} movements are analyzed to detect if a visible person is speaking. However, the performance is limited due to weak correlation between the body motion and speaking activities. Besides, non-speaking activities, e.g., licking lips, eating food and grinning, may also degrade the ASD performance. Despite these limitations, the audio or visual single modal solutions serve as the foundation for ASD.

%\textcolor{blue}{I introduce two types of ASD approaches here. Assignment vs Classification}
% Methods for Audio-visual ASD. How they did and what we learn.

Audio-visual processing has seen significant benefits through modality fusion~\cite{minotto2015multimodal,ephrat2018looking}. As the speech rhythm and word pronunciation are closely correlated with facial motion, an interesting and promising alternative is to combine both audio and vision information to perform ASD. Exploring audio-visual ASD, one approach is to view it as an assignment task. It is assumed that the detected speech must belong to one of the speakers on the screen~\cite{alcazar2020active,leon2021maas}. However, this assumption does not always hold because there could be cross talk or off-screen speech in practice. Another approach is to perform ASD as a classification task to evaluate the visible face on the screen one-by-one. Some studies~\cite{chung2016out, chung2019naver, afouras2020self} simply concatenate the extracted audio and visual features as the input, and apply a multi-layer perceptron (MLP)-based binary classifier to detect the active speaker at each short video segment, without considering the inter-frame temporal dependency. Others further adopt the backend classifier with temporal structure like recurrent neural network (RNN)~\cite{tao2017bimodal, tao2019end}, gated recurrent unit (GRU)~\cite{roth2020ava} and long short-term memory (LSTM)~\cite{zhangmulti, sharma2020crossmodal, ariav2019end}, which have achieved preliminary success. Our proposed TalkNet is motivated by this thought. 

% Long-term audio and visual feature learning is better
%\textcolor{blue}{To Prof Li: Please check this paragraph and the section 2.2. We once modified it a few days ago.}

\begin{figure*}[t]
  \centering
  \includegraphics[width=\linewidth]{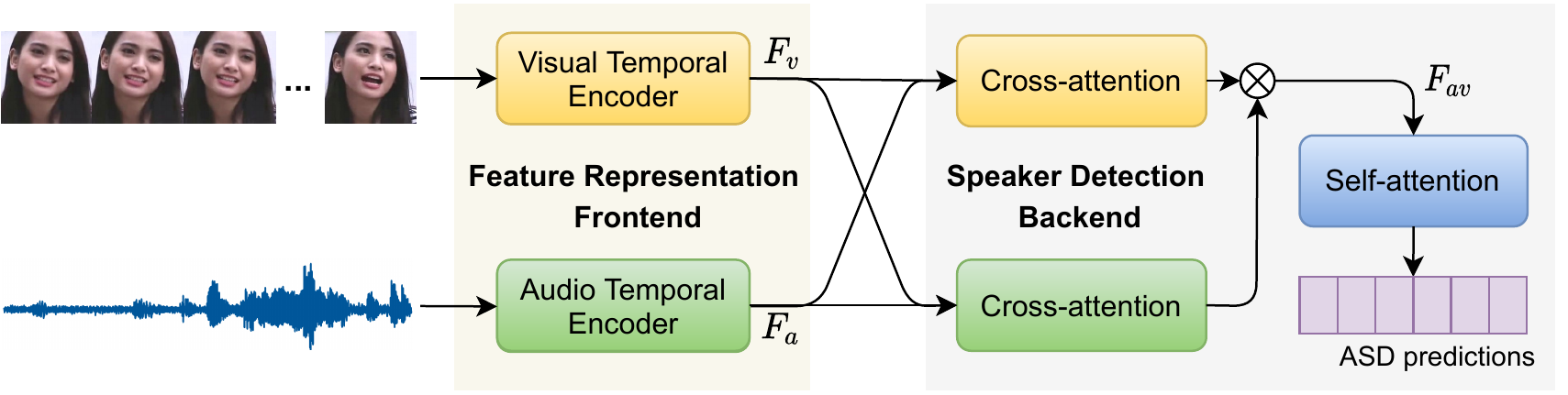}
  \caption{An overview of our TalkNet, which consists of visual and audio temporal encoders followed by cross-attention and self-attention for ASD prediction.}
  \label{fig:TalkNet detail}
  \Description{TalkNet}
\end{figure*}

\subsection{Long-term Temporal Context}

As ASD seeks to make a fine-grained decision at audio or video frame level, most of the prior studies employ short-term features and make decisions at split segments of less than 600 ms. While smoothing method can be used to aggregate short-term decisions for for long-term video~\cite{chung2019naver, alcazar2020active}, the potential of long-term features has not been fully explored yet.

It is common that ASD uses individual uni-modal frontend feature extractors to learn the audio and visual embeddings, that is followed by the backend classifier to incorporate audio-visual synchrony. For the uni-modal representation learning, the utterance-level model performs better than the frame-level model in the audio tasks such as audio classification~\cite{pons2019randomly, fernandes2021detecting}. The recent studies in video object detection also show that it is beneficial to leverage the temporal context at the proposal-level by end-to-end optimization to learn the completed video presentations~\cite{shvets2019leveraging, wu2019sequence}. As their short-term embeddings encode long-term temporal context, such techniques generally provide improved performance when making short-term decisions. The success in these studies motivates us to consider encoding long-term audio and visual temporal context at the utterance level for ASD task. 

% Long-term audio-visual synchronization learning is better
On the other hand, audio-visual ASD takes advantage of the cross-modal synchronization information. In audio-visual synchronization studies, by using convolutional neural network (CNN)~\cite{owens2018audio, chung2019perfect}, LSTM~\cite{shalev2020end} or attention model~\cite{cheng2020look}, the longer video utterance are used, the more representative features can be extracted, which eventually boost the performance~\cite{chung2016out, chung2019perfect}. These studies demonstrate that long-term temporal context is significantly important to learn the audio-visual relationship in ASD. As ASD aims to learn the modality feature and the audio-visual relationship, we believe it will benefit from long-term temporal context either from intra-modal signals or inter-modal signals.

\section{TalkNet}
\label{seciii}

TalkNet is an end-to-end pipeline that takes the cropped face video and corresponding audio as input, and decide if the person is speaking in each video frame. It consists of a feature representation frontend, and a speaker detection backend classifier, as illustrated in Figure~\ref{fig:TalkNet detail}. The frontend contains an audio temporal encoder and a video temporal encoder. They encode the frame-based input audio and video signals into the time sequence of audio and video embeddings, that represent temporal context. The backend classifier consists of an inter-modality cross-attention mechanism to dynamically align audio and visual content, and a self-attention mechanism to observe speaking activities from the temporal context at the utterance level.

\subsection{Visual Temporal Encoder}

\begin{figure}[b]
  \centering
  \includegraphics[width=0.9\linewidth]{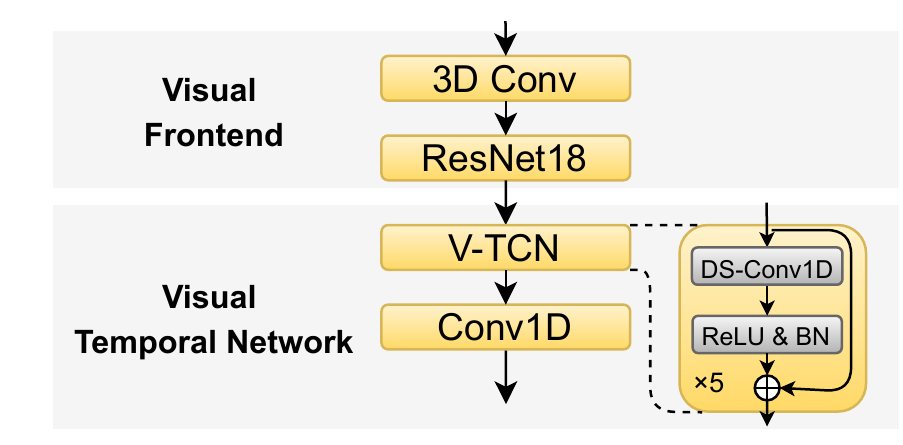}
  \caption{The structure of visual temporal encoder, which contains the visual frontend and the visual temporal network. $\bigoplus$ denotes point-wise addition.}
  \label{fig:visual encoder}
\end{figure}

The visual temporal encoder aims to learn the long-term representation of facial expression dynamics. As illustrated in Figure~\ref{fig:visual encoder}, it consists of the visual frontend and the visual temporal network. We seek to encode the visual stream into a sequence of visual embeddings $F_v$ that have the same time resolution.

The visual frontend explores spatial information within each video frame. It consists of a 3D convolutional layer (3D Conv) followed by a ResNet18 block~\cite{afouras2018conversation}. This frontend encodes the video frame stream into a sequence of frame-based embedding. The visual temporal network consists of a video temporal convolutional block (V-TCN), which has five residual connected rectified linear unit (ReLU), batch normalization (BN) and depth-wise separable convolutional layers (DS Conv1D)~\cite{afouras2018deep}, followed by a Conv1D layer to reduce the feature dimension. It aims to represent the temporal content in a long-term visual spatio-temporal structure. For example, for a visual temporal encoder that has a receptive field of 21 video frames, we take a segment of up to 840 ms to encode a video embedding when the video frame rate is 25 frame-per-second (fps).

\subsection{Audio Temporal Encoder}
The audio temporal encoder seeks to learn an audio content representation from the temporal dynamics. It is a 2D ResNet34 network with squeeze-and-excitation (SE) module~\cite{hu2018squeeze} introduced in~\cite{chung2020defence}. An audio frame is first represented by a vector of Mel-frequency cepstral coefficients (MFCCs). The audio temporal encoder takes the sequence of audio frames as the input, generate the sequence of audio embeddings $F_a$ as the output. The ResNet34 are designed with dilated convolutions such that the time resolution of audio embeddings $F_a$ matches that of the visual embeddings $F_v$ to facilitate subsequent attention mechanism. {For example, the audio temporal encoder has a receptive field of 189 audio frames. In other words, we take a segment of 1,890 ms to encode an audio embedding, when the MFCC window step is 10 ms, to capture the long-term temporal context.}

\begin{figure}
    \centering
    \subfloat[\centering ]{{\includegraphics[width=0.63\linewidth]{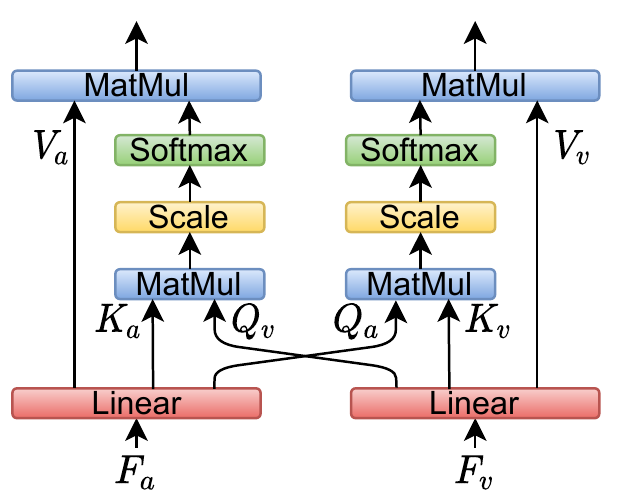}}}
    %\qquad
    \subfloat[\centering ]{{\includegraphics[width=0.34\linewidth]{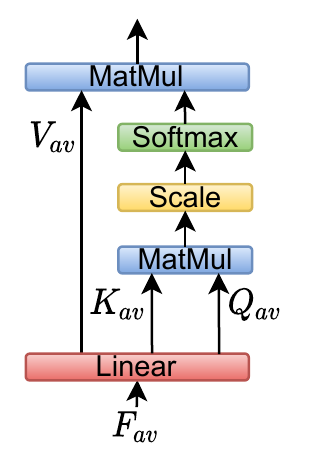}}}
    \caption{(a) The attention layer in the cross-attention network. Considering the audio embeddings $F_{a}$ as the source, and the visual feature $F_{v}$ as the target, we generate audio attention feature $F_{a \rightarrow v}$ as the output. Similarly, we generate visual attention feature $F_{v \rightarrow a}$. (b) The attention layer in the self-attention network.}
    \label{fig:cross}
\end{figure}

\subsection{Audio-visual Cross-Attention}
$F_a$ and $F_v$ are expected to characterize the events that are relevant to speaking activities for audio and visual, respectively. We are motivated by the fact that audio-visual synchronization is an informative cue for speaking activities as well. As audio and visual flow each has its own dynamics, they are not exactly time aligned. The actual audio-visual alignment may depend on the instantaneous phonetic content and the speaking behavior of the speakers. We propose two cross-attention networks along the temporal dimension to dynamically describe such audio-visual interaction.

The core part of the cross-attention network is the attention layer, which is shown in Figure~\ref{fig:cross} (a). The inputs are the vectors of query ($Q_a,Q_v$), key ($K_a,K_v$), and value ($V_a,V_v$) from audio and visual embeddings, respectively, projected by a linear layer. The outputs are the audio attention feature $F_{a \rightarrow v}$ and visual attention feature $F_{v \rightarrow a}$ as formulated in Eq. (\ref{e1}) and Eq. (\ref{e2}), where $d$ denotes the dimension of $Q$, $K$ and $V$.
\begin{equation}
    \label{e1}
    F_{a \rightarrow v} = softmax(\frac{{Q_v}{K_a^T}}{\sqrt{d}})V_a
\end{equation}
\begin{equation}
    \label{e2}
    F_{v \rightarrow a} = softmax(\frac{{Q_a}{K_v^T}}{\sqrt{d}})V_v
\end{equation}
As formulated in Eq. (\ref{e1}) and Eq. (\ref{e2}), to learn the interacted new audio feature $F_{a \rightarrow v}$, the attention layer applies $F_v$ as the target sequence to generate query, and $F_a$ as the source sequence to generate key and value, and to learn $F_{v \rightarrow a}$ vice versa. The attention layer is followed by the feed-forward layer. Residual connection and layer normalization are also applied after these two layers to generate the whole cross-modal attention network. The outputs are concatenated together along the temporal direction.

\subsection{Self-Attention and Classifier}

A self-attention network is applied after the cross-attention network to model the audio-visual utterance-level temporal information. As illustrated in Figure~\ref{fig:cross} (b), this network is similar to the cross-attention network except that now the query ($Q_{av}$), key ($K_{av}$) and value ($V_{av}$) in the attention layer all come from the joint audio-visual feature $F_{av}$. With the self-attention layer, we seek to distinguish the speaking and non-speaking frames.

\subsection{Loss Function}
We finally apply a fully connected layer followed by a softmax operation to project the output of the self-attention network to an ASD label sequence. We view ASD as a frame-level classification task. The predicted label sequence is compared with the ground truth label sequence by cross-entropy loss. The loss function is presented in Eq (\ref{formula:loss}), where $s_i$ and $y_i$ are the predicted and the ground truth ASD labels of $i^{{\text {th}}}$ video frame, $i \in [1,T]$. $T$ refers to the number of video frames.

\begin{equation}
    Loss = - \frac{1}{{T}} \sum_{i=1}^{{T}} (y_i \cdot \mathrm{log}\; s_i + (1-y_i) \cdot \mathrm{log}\; (1-s_i))
    \label{formula:loss}
\end{equation}

\subsection{Audio Augmentation with Negative Sampling}
The noise-robustness of ASD in the presence of noise and interference speakers remains a challenging topic. One traditional audio augmentation method is to use a large noise dataset~\cite{xvectors} to augment the training data by overlaying the noise on top of the original sound track. This method involves the external data source to increase the diversity. However, it is not straightforward to find such acoustic data that matches the video scenes. 

To increase the amount of samples, we propose a negative sampling method to offer a simple yet effective solution. In practice, we use one video as the input data during training, and then we randomly select the audio track from another video in the same batch as the noise to perform audio augmentation. Such augmented data effectively have the same label, e.g., active speaker or inactive speaker, as the original sound track. This approach involves the in-domain noise and interference speakers from the training set itself. It does not require data outside the training set for audio augmentation. 

\section{Experiments}
\label{seciv}
\subsection{Dataset}
\subsubsection{\textbf{AVA-ActiveSpeaker}}
% Statistics of AVA
The AVA-ActiveSpeaker dataset\footnote{\url{https://research.google.com/ava/download.html##ava_active_speaker_download}} is derived from Hollywood movies~\cite{roth2020ava}. It contains $29,723$, $8,015$ and $21,361$ video utterances in the training, validation and test sets, respectively. The video utterances range from 1 to 10 seconds and are provided as face tracks. We follow the official evaluation tool and report the performance in terms of mean average precision (mAP). 

There are several challenges involved in the AVA-ActiveSpeaker dataset.
% AVA-ActiveSpeaker dataset presents several challenges in the evaluation. 
The language is diverse, and the frame per second (fps) of the movies varies. Furthermore, a significant number of videos have blurry images and noisy audio. It also contains many old movies with dubbed dialogues.
All these factors make it hard to accurately synchronize the audio-visual signals.

\subsubsection{\textbf{Columbia Active Speaker Dataset}}
The Columbia ASD dataset\footnote{\url{http://www.jaychakravarty.com/active-speaker-detection/}} is a standard benchmark test dataset for ASD~\cite{chakravarty2016cross}. It contains an 87-minute panel discussion video, with 5 speakers taking turns to speak, in which 2-3 speakers are visible at any given time. We follow the common protocol of this benchmark to use F1 score as the evaluation metric. The Columbia ASD dataset doesn't provide a common splitting between training and test sets.  

\subsubsection{\textbf{TalkSet}}
\label{talkset}
Due to its limited size, the Columbia ASD dataset is usually only used as a test set. Furthermore, the AVA-ActiveSpeaker dataset is labelled with face bounding boxes with a different algorithm, which are incompatible with those of the Columbia ASD dataset. We are motivated by the call for an audio-visual ASD dataset that covers real-world scenarios. This leads to the idea of a new database. We leverage two large-scale audio-visual datasets in the wild, LRS3~\cite{afouras2018lrs3} and VoxCeleb2~\cite{chung2018voxceleb2}, to form a new ASD dataset, named ``TalkSet'', that covers all valid ASD conditions.
 
First, we consider that humans detect active speakers by examining three aspects of a video, 1) On audio signal, is there an active voice? 2) For visual signal, are the lips of someone moving? 3) When there is an active voice and the lips of someone are moving, is the voice synchronized with the lips movement? The above three cues lead to five valid conditions in the real world, which are summarized in Table~\ref{tab:ASD conditions}.

We select 90,000 videos with active voice from VoxCeleb2\footnote{\url{https://www.robots.ox.ac.uk/~vgg/data/voxceleb/vox2.html}}~\cite{chung2018voxceleb2}. We also collect 60,000 videos without an active voice, at the same time, longer than one second from LRS3\footnote{\url{https://www.robots.ox.ac.uk/~vgg/data/lip_reading/lrs3.html}}~\cite{afouras2018lrs3} using the Kaldi-based voice activity detection system~\cite{povey2011kaldi}. In total, we have got 150,000 videos that range from 1 to 6 seconds. The total length of these videos is 151.65 hours, out of which 71.45 hours are speaking and 80.20 hours are non-speaking. We randomly split it into 135,000 videos for training and 15,000 videos for validation. Finally, we adopt the Columbia ASD dataset as the test data. 

We understand that both LRS3 and VoxCeleb2 use the S3FD face detection method~\cite{zhang2017s3fd} to provide ground truth face tracking of the speakers. To be consistently, we also apply the same method for face tracking on unknown test videos, including the Columbia ASD dataset.

\begin{table}
  \caption{All valid conditions of ASD videos in the wild (Note: only when the audio is active and the lips are moving do we consider whether audio-visual is synchronized.)}
  \label{tab:ASD conditions}
  \begin{tabular}{p{0.7cm}<{\centering}p{1.3cm}<{\centering}p{1.6cm}<{\centering}p{1.3cm}<{\centering}p{1.8cm}<{\centering}}
    \hline
    \textbf{Index} & \textbf{Audio} & \textbf{Lips} & \textbf{Audio-visual} & \textbf{ASD Label}\\
    \hline
    1 & active & moving & sync & speaking \\
    2 & active & moving & not sync & non-speaking \\
    3 & active & not moving & NA & non-speaking \\
    4 & inactive & moving & NA & non-speaking \\
    5 & inactive & not moving & NA & non-speaking \\
    \hline
  \end{tabular}
\end{table}	

\subsection{Implementation Details}

We build the TalkNet using the PyTorch library with the Adam optimizer. The initial learning rate is $10^{-4}$, and we decrease it by $5\%$ for every epoch. The dimension of MFCC is $13$. All the faces are reshaped into $112 \times 112$. We set the dimensions of the audio and visual feature as 128. Both cross-attention and self-attention network contain one transformer layer with eight attention heads. 

We randomly flip, rotate and crop the original images to perform visual augmentation. As the Columbia ASD dataset is an open dataset, we apply the additional sources from RIRs data~\cite{RIRS} and the MUSAN dataset~\cite{MUSAN} to perform audio augmentation on the TalkSet during training, and evaluate the performance using Sklearn
library\footnote{\url{https://scikit-learn.org/stable/modules/generated/sklearn.metrics.f1_score.html}}. For the AVA-ActiveSpeaker dataset, we apply the proposed negative sampling technique to add the in-domain noise from the training set itself. Finally, we evaluate the performance on the test set using the official tool\footnote{\url{https://github.com/activitynet/ActivityNet}}. We also evaluate the performance on the validation set as it comes with the ground truth labels for quick examination.

\section{Results}
\label{secv}
\subsection{Comparison with the State-of-the-art}
We now compare the proposed TalkNet with the state-of-the-art systems on both the AVA-ActiveSpeaker and Columbia ASD dataset. 
% AVA
First, we summarize the results on the AVA-ActiveSpeaker dataset in Table~\ref{tab:AVA Val}. We observe that TalkNet achieves $92.3\%$ mAP and outperforms the best competitive system, i.e., MAAS-TAN~\cite{leon2021maas}, by $3.5\%$ on the validation set. Some studies report their results in terms of Area under the Curve of ROC (AUC) on the same validation set. For ease of comparison, we also report the comparison of AUC results in Table~\ref{tab:AVA Val AUC}. Without surprise, the TalkNet also achieves $3.6\%$ improvement over the best reported AUC, cf., Huang et al.~\cite{huang2020improved}. 
\begin{table}[htb]
  \caption{Comparison with the state-of-the-art on the AVA-ActiveSpeaker validation set in terms of mean average precision (mAP).}
  
  \label{tab:AVA Val}
  \begin{tabular}{p{4cm}<{\centering}p{2cm}<{\centering}}
    \hline
   % \multicolumn{2}{c}{\textbf{AVA-ActiveSpeaker Validation Set}}\\
    \textbf{Method} & \textbf{mAP (\%)}\\
    \hline
    Roth et al.~\cite{roth2020ava, leon2021maas} & $79.2$ \\
    Zhang et al.~\cite{zhangmulti}& $84.0$\\
	MAAS-LAN~\cite{leon2021maas}& $85.1$\\
	Alcazar et al.~\cite{alcazar2020active} & $87.1$\\
	Chung et al~\cite{chung2019naver} & $87.8$\\
	MAAS-TAN~\cite{leon2021maas} & $88.8$\\
    \textbf{TalkNet (proposed)} & $\textbf{92.3}$\\
    \hline
  \end{tabular}
\end{table}

\begin{table}[h]
  \caption{Comparison with the state-of-the-art on the AVA-ActiveSpeaker validation set in terms of area under the curve (AUC).}
  \label{tab:AVA Val AUC}
    \begin{tabular}{p{4cm}<{\centering}p{2cm}<{\centering}}
    \hline
    \textbf{Model} & \textbf{AUC (\%)}\\
    \hline
    Sharma et al.~\cite{sharma2020crossmodal} & $82.0$\\
    Roth et al.~\cite{roth2020ava} & $92.0$\\
    Huang et al.~\cite{huang2020improved}& $93.2$\\
    \textbf{TalkNet (proposed)} & $\textbf{96.8}$\\
    \hline
  \end{tabular}
\end{table}
As the ground truth labels of the AVA-ActiveSpeaker test set are not available to the public, we obtain the evaluation results in Table~\ref{tab:AVA Test} on the test set with the assistance of the organizer. Our $90.8\%$ mAP also outperforms the best prior work by $3.0\%$, cf., Chung et al.~\cite{chung2019naver}. 

Note that some prior studies~\cite{leon2021maas, alcazar2020active} applied additional networks to learn the relationship among the cropped face videos. Others~\cite{chung2019naver, zhangmulti} used the pre-trained model in another large-scale dataset. By contrast, TalkNet only uses the AVA-ActiveSpeaker training set to train the single face videos from scratch without any additional post-processing. We believe that pre-training and other advanced techniques will further improve TalkNet, which is beyond the scope of this paper.

\begin{table}[htb]
  \caption{Comparison with the state-of-the-art on the AVA-ActiveSpeaker test set in terms of mAP.}  
  \label{tab:AVA Test}
  \begin{tabular}{p{4cm}<{\centering}p{2cm}<{\centering}}
    \hline
    \textbf{Method} & \textbf{mAP (\%)}\\
    \hline
    Roth et al.~\cite{roth2020ava} & $82.1$\\
    Zhang et al.~\cite{zhangmulti}& $83.5$\\
    Alcazar et al.~\cite{alcazar2020active} & $86.7$\\
    Chung et al.~\cite{chung2019naver} & $87.8$\\
    \textbf{TalkNet (proposed)} & $\textbf{90.8}$\\
    \hline
  \end{tabular}
\end{table}

% Columbia ASD
We then evaluate TalkNet on the Columbia Active Speaker Detection dataset. Its performance along with comparison to other existing methods are shown in Table~\ref{tab:TalkNet}. We observe that the F1 score, which is the standard metric in this benchmark, is the maximum for proposed TalkNet, which is $96.2\%$ for the average result that has an improvement over the best existing system by $2.2\%$. For all the five speakers, TalkNet provides the best performance for three of them (Bell, Lieb and Sick). It is noted that Columbia ASD is an open-training dataset, so the methods in Table~\ref{tab:TalkNet} are trained on different data, so we only claim that our TalkNet is efficient on the Columbia ASD dataset.
\begin{table}[h]
  \caption{Comparison with the state-of-the-art on the Columbia ASD dataset in terms of F1 scores (\%).}
  \label{tab:TalkNet}
  \begin{tabular}{p{3cm}<{\centering}p{0.5cm}<{\centering}p{0.5cm}<{\centering}p{0.5cm}<{\centering}p{0.5cm}<{\centering}p{0.5cm}<{\centering}|p{0.5cm}<{\centering}}
    \hline
    \multirow{2}{*}{\textbf{Method}} & \multicolumn{6}{c}{\textbf{Speaker}}\\
    {} & \textbf{Bell} & \textbf{Boll} & \textbf{Lieb} & \textbf{Long} & \textbf{Sick} & \textbf{Avg.}\\
    \hline
    Brox et al.~\cite{brox2004high, shahid2019comparisons} & $84.1$ & $72.3$ & $80.6$ & $60.0$ & $68.9$ & $73.2$~~~\\
    Chakravarty et al.~\cite{chakravarty2016cross} & $82.9$ & $65.8$ & $73.6$ & $86.9$ & $81.8$ & $78.2$~~~\\
    Zach et al.~\cite{zach2007duality, shahid2019comparisons} & $89.2$ & $88.8$ & $85.8$ & $81.4$ & $86.0$ & $86.2$~~~\\
    RGB-DI~\cite{shahid2019comparisons} & $86.3$ & ${93.8}$ & $92.3$ & $76.1$ & $86.3$ & $87.0$~~~\\
    SyncNet~\cite{chung2016out} & $93.7$ & $83.4$ & $86.8$ & ${\textbf{97.7}}$ & $86.1$ & $89.5$~~~\\
    LWTNet~\cite{afouras2020self} & $92.6$ & $82.4$ & $88.7$ & $94.4$ & $95.9$ & $90.8$~~~\\
    RealVAD~\cite{beyan2020realvad} & $92.0$ & $\textbf{98.9}$ & $94.1$ & $89.1$ & $92.8$ & $93.4$~~~\\
    S-VVAD~\cite{shahid2021s} & $92.4$ & $97.2$ & $92.3$ & $95.5$ & $92.5$ & $94.0$~~~\\
    \textbf{TalkNet (proposed)}     & $\textbf{{97.1}}$ & $90.0$ & ${\textbf{99.1}}$ & ${96.6}$ & ${\textbf{98.1}}$ & $\textbf{96.2}$~~~\\
    \hline
  \end{tabular}
\end{table}

\subsection{Ablation Study}

\begin{figure*}[h]
  \centering
  \includegraphics[width=1\linewidth]{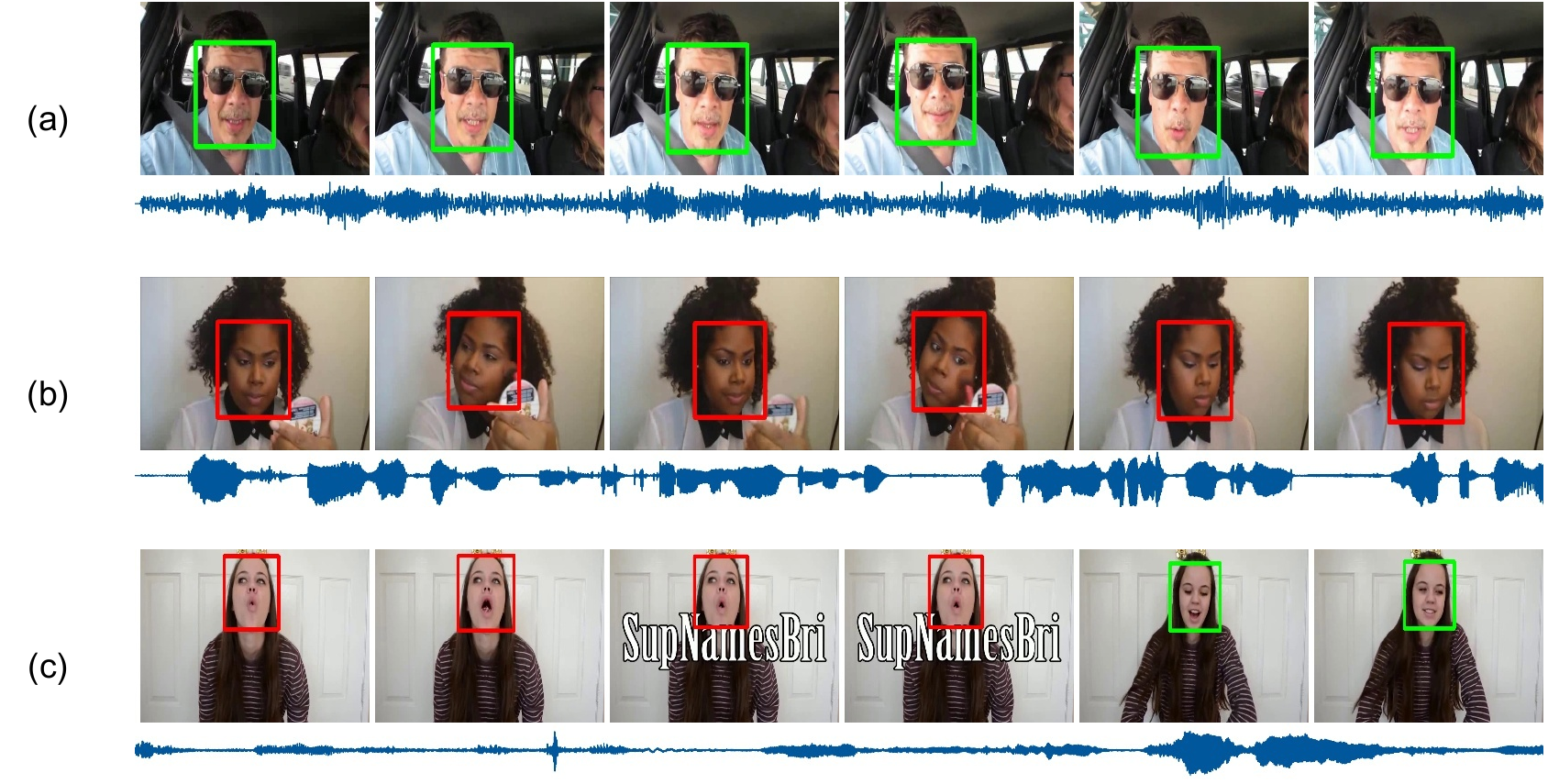}
  \caption{Results of TalkNet for the real-world videos with one person on the screen. The green box denotes the active speaker. The red box denotes the inactive speaker. The time interval between the adjacent images is 1 second. (a) The man is speaking in the noisy environment.  (b) The woman is introducing the makeup process through dubbing. So the speech is not synchronized with her lip movement. (c) The woman is eating candy. Although her lips are always moving, she is not speaking in the beginning.}
  \label{fig:visulization}
\end{figure*}
\begin{figure*}[h]
  \centering
  \includegraphics[width=1\linewidth]{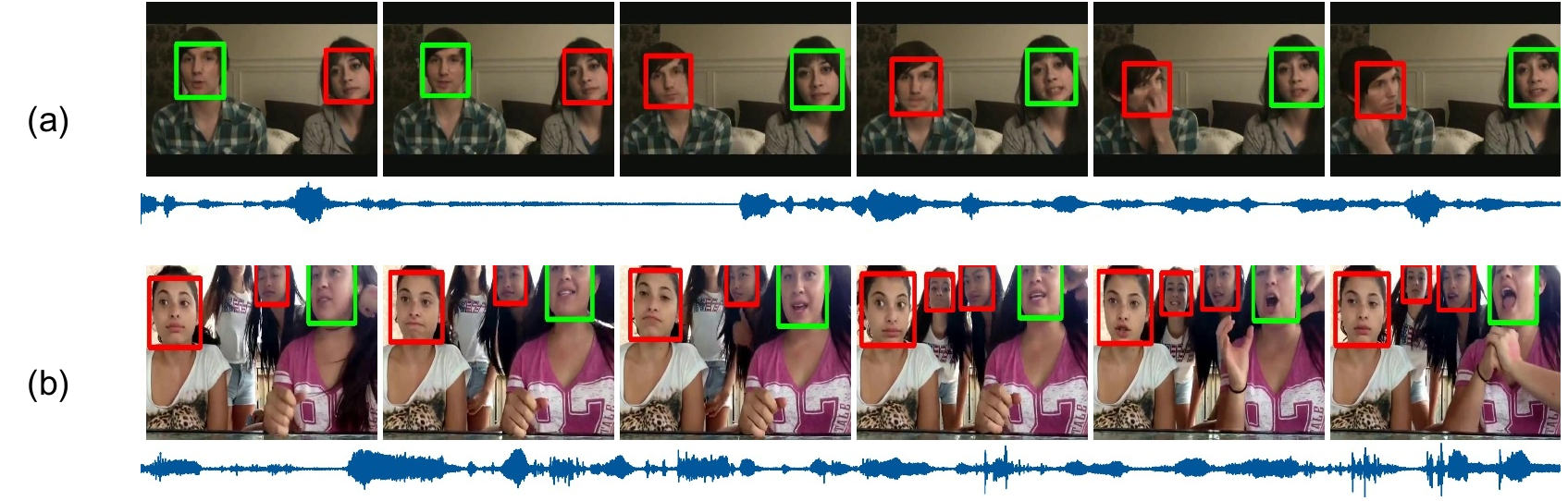}
  \caption{Results of TalkNet for the real-world videos with multiple persons on the screen. (a) Two speakers take turns speaking, and the man's lips are concealed sometimes. (b) Four speakers are talking in a boisterous environment with background music. Everyone's mouth is moving, but only the girl on the right is speaking.}
  \label{fig:visual_multi}
\end{figure*}

We further perform experiments to analyze the contributions of individual techniques deployed in TalkNet.

\subsubsection{\textbf{Long-term sequence-level temporal context}}
The prior studies usually use short-term features of 5 to 13 video frames on the AVA-Activespeaker dataset~\cite{chung2019naver, leon2021maas, alcazar2020active} for video embedding. We believe that long-term features are more evident of speaking episode. To study the difference between long-term and short-term features, we use a fixed number of $N$ frames instead of the entire video sequence during training and testing, where $N$ is chosen from {5,10,25,50,100} that amounts to  0.2, 0.4, 1, 2 and 4 second. 

We report the evaluation results of TalkNet in Table~\ref{tab:AVA num_of_frames}, and observe that the system can hardly work with very short video segment, e.g., when $N=5$, as there is not enough temporal context in a 0.2-second segment. As the duration of video increases, mAP improves consistently from 75.2\% to 89.4\%. 

As we increase the duration of videos, there are a fewer number video segments for training. As a result, we don't observe improvement from 50 frames to 100 frames of video duration. This study confirms our hypothesis that the long-term sequence-level information is a major source of contributions to the improved performance.  

\begin{table}[htb]
  \caption{Performance evaluation by the length of the video on the AVA-ActiveSpeaker validation set. We use a fixed number of video frames during both training and testing.}
  \label{tab:AVA num_of_frames}
  \begin{tabular}{p{2.5cm}<{\centering}p{2.5cm}<{\centering}p{2.5cm}<{\centering}}
    \hline
    \multirow{2}{*}{\textbf{\# video frames}} & \textbf{Length (APRX seconds)} & \multirow{2}{*}{\textbf{mAP(\%)}}\\
    \hline
    5 & 0.2 & ${75.2}$ \\
    10 & 0.4 &${82.8}$ \\
    25 & 1 &${87.9}$ \\
    50 & 2 &${89.5}$ \\
    100 & 4 &${89.4}$ \\
    \textbf{Variable} & \textbf{1 - 10}  &$\textbf{92.3}$ \\
    \hline
  \end{tabular}
\end{table}	

\subsubsection{\textbf{Short-term vs long-term features}} 
To appreciate the contribution of the long-term features, we further compare TalkNet with a prior work~\cite{alcazar2020active}, which uses the short-term audio and visual embedding and the relationship between the co-occurring speakers via a two-step training. The first step encodes the low-dimensional representation for video segments of 440 ms, and fuses the audio-visual information, which is similar to TalkNet except that there is neither temporal encoder to increase the receptive fields, and nor attention mechanism.  
 
We first reproduce the system in~\cite{alcazar2020active} to obtain 78.2\% mAP for 11 video frames input, which is slightly lower than 79.5\% mAP in the original paper due to different batch size setting. Then we extend the segments to 25 video frames and compare the results in Table~\ref{tab:ASC compare}. 
We observe that the TalkNet obtains a 4.8\% improvement from the longer input videos. However, the prior work ~\cite{alcazar2020active} does not benefit from the longer segments, with a performance drop of  2.1\%. This study suggests that longer video duration doesn't help without the long audio and visual receptive fields and an adequate attention mechanism.

\begin{table}[H]
  \caption{A contrastive study between two systems on efficient use of long video segments on the AVA-ActiveSpeaker validation set in terms of mAP (\%).}
  \label{tab:ASC compare}
  \begin{tabular}{p{2.5cm}<{\centering}p{1cm}<{\centering}p{1cm}<{\centering}p{2cm}<{\centering}}
    \hline
    \multirow{2}{*}{\textbf{Method}} & \multicolumn{2}{c}{\textbf{\# video frames}} & \multirow{2}{*}{\textbf{Change}}  \\
    & \textbf{\textbf{11}} & \textbf{\textbf{25}} & \\
    \hline
    Alcazar et al.~\cite{alcazar2020active} & ${78.2}$ & ${76.1}$ & ${-2.1}$\\
    TalkNet (Proposed) & ${83.1}$ & ${87.9}$ & ${+4.8}$ \\
    \hline
  \end{tabular}
\end{table}

\subsubsection{\textbf{Ablation study of TalkNet attention mechanism}}
To show the contribution of audio-visual relationship and interaction over a long-term video to the ASD task, we conduct the ablation study for the audio-visual attention network in TalkNet. The results are summarized in Table~\ref{tab:AVA interaction}. We find that, without cross-attention or self-attention, the performance will drop 0.7\% or 1.4\% mAP, respectively on the AVA-ActiveSpeaker validation set. When removing the whole audio-visual attention network, the result will decrease to only 90.0\% mAP by 2.3\%. The results confirm the effectiveness of the cross-attention and self-attention in learning inter-modality cues and long-term audio-visual temporal context.  

\begin{table}[htb]
  \caption{Ablation study of the cross-attention and self-attention mechanisms in TalkNet on the AVA-ActiveSpeaker validation set.}
  \label{tab:AVA interaction}
  \begin{tabular}{p{4cm}<{\centering}p{2cm}<{\centering}}
    \hline
    \textbf{Model} & \textbf{mAP(\%)} \\
    \hline
    w/o Both & $90.0$ \\
    w/o Self-attention & $90.9$ \\
    w/o Cross-attention& $91.6$ \\
    \textbf{TalkNet} & $\textbf{92.3}$ \\
    \hline
  \end{tabular}
\end{table}	

\subsubsection{\textbf{Audio augmentation}} 
We report the audio augmentation experiments of TalkNet on the AVA-ActiveSpeaker validation set in Table~\ref{tab:AVA aug}. `With neg\_sampling', `With noise\_aug' and `W/o audio\_aug' stand for our proposed negative sampling method, the traditional audio augmentation method which involves a large noise dataset, and without any audio data augmentation. The proposed TalkNet without any audio data augmentation still outperforms the state-of-the-art. We also observe that there is an obvious difference with and without negative sampling technique. The proposed negative sampling technique outperforms the traditional audio augmentation. These results confirm the efficiency of the negative sampling method, which doesn't involve external data.

\begin{table}[htb]
  \caption{Evaluation of TalkNet with and without audio data augmentation on the AVA-ActiveSpeaker validation set.}
  \label{tab:AVA aug}
  \begin{tabular}{p{4cm}<{\centering}p{2cm}<{\centering}}
    \hline
    \textbf{Augmentation conditions} & \textbf{mAP(\%)} \\
    \hline
    {W/o audio\_aug} & ${89.4}$ \\
    {With noise\_aug} & ${92.2}$\\
    {\textbf{With neg\_sampling}} & $\textbf{{92.3}}$ \\
    \hline
  \end{tabular}
\end{table}	

\subsection{Qualitative Analysis}

\begin{figure}[htb]
    \centering
    \subfloat[\centering ]{{\includegraphics[width=\linewidth]{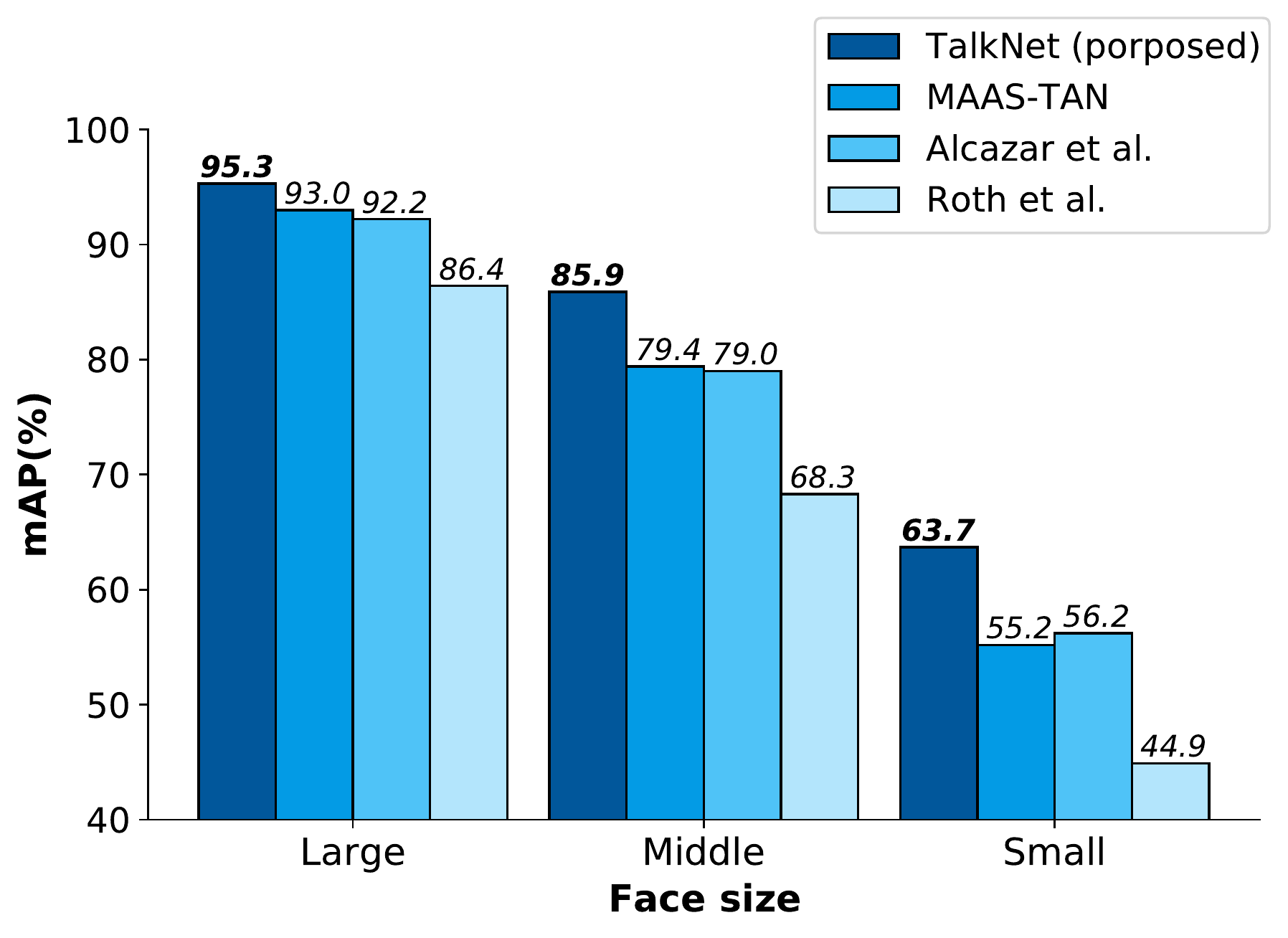}}}%
    \qquad
    \subfloat[\centering ]{{\includegraphics[width=\linewidth]{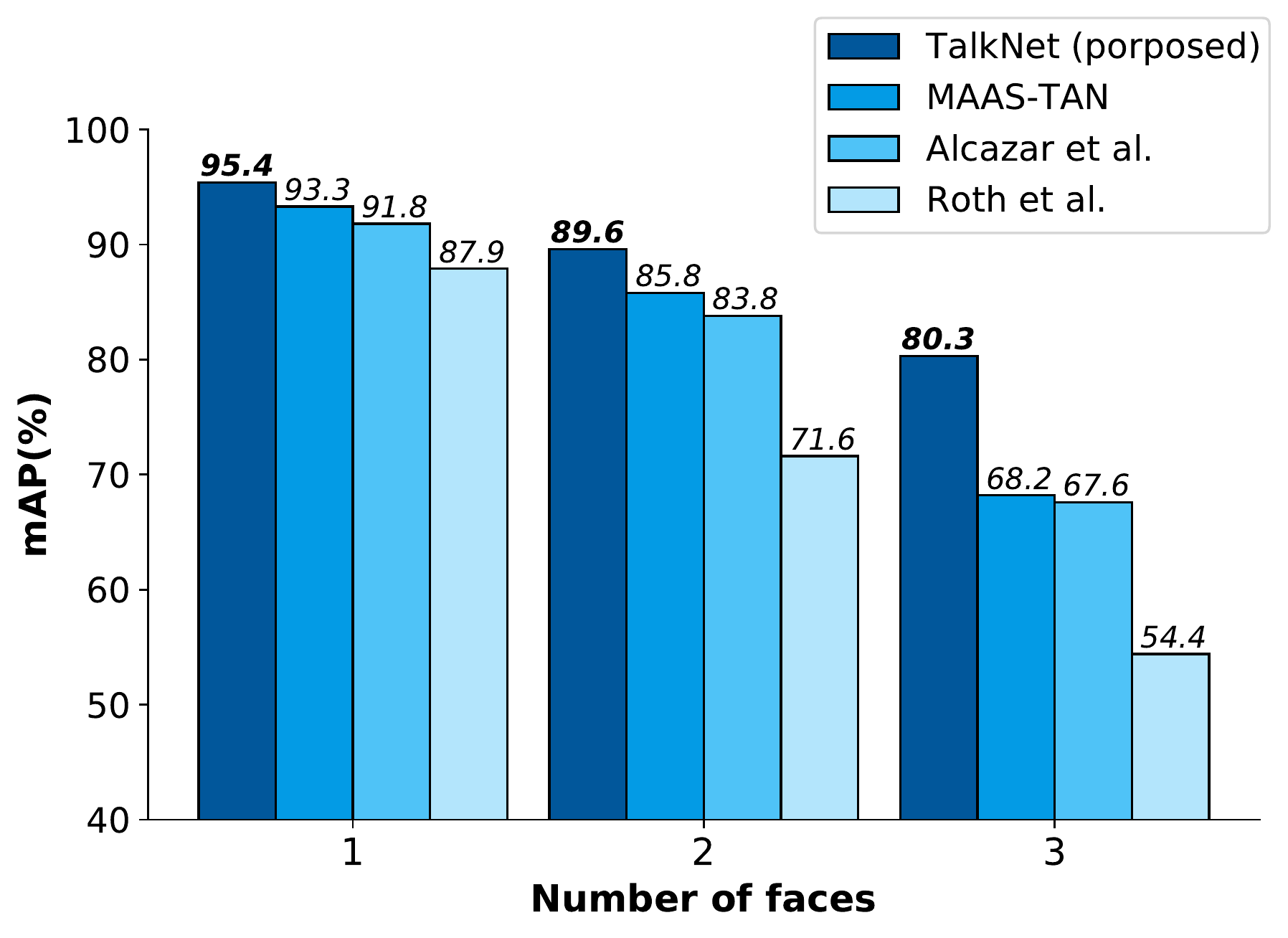}}}%
    \caption{The performance of our TalkNet and other competing methods for (a) various face sizes, and (b) specific face size in the same video frame.}%
    \label{fig:analysis}%
\end{figure}
Figure~\ref{fig:visulization} and Figure~\ref{fig:visual_multi} show some results of TalkNet in the real-world videos with one person and multiple persons on the screen, respectively. 

In Figure~\ref{fig:analysis}(a), we report the performance for different face sizes on the AVA-ActiveSpeaker validation set. `Small': The face width is smaller than 64 pixels; `Middle': The face width is between 64 to 128 pixels; `Large': The face width is larger than 128 pixels. We observe that the performance decreases as face size gets smaller.  It is worth noting that our proposed TalkNet always achieves the best results across all face sizes.

In Figure~\ref{fig:analysis}(b), we study the effect of the number of visible faces in a video frame, i.e.,  1, 2 or 3 faces, which represent about 90\% of all the validation data. From Figure~\ref{fig:analysis}(b), we observe that the ASD task becomes more challenging as the number of faces increases. While the performance drops across all methods, the proposed TalkNet is clearly more robust than other competing methods. 

% We attribute the robustness of TalkNet to two mechanisms, one is the long-term features that are considered to be more robust against noise, another is the negative sampling technique for audio augmentation. The results also suggest that TalkNet performs well in both favourable and unseen adverse environments.

\section{Conclusion}
\label{conc}
In this work, we address and study audio-visual ASD with long-term temporal features. TalkNet utilizes the sentence-level audio-visual videos information as the input to explore the audio-visual relationship and synchronization information. TalkNet outperforms the state-of-the-art results in both two mainstream ASD benchmark by 3.5\% and 2.2\%. 

\section{Acknowledgment}
This work was supported by Grant No. A18A2b0046 from the Singapore Government’s Research, Innovation and Enterprise 2020 plan (Advanced Manufacturing and Engineering domain); Grant No. 1922500054 from the Science and Engineering Research Council, Agency of Science, Technology and Research, Singapore, the National Robotics Program.

\newpage
\bibliographystyle{ACM-Reference-Format}
\bibliography{ref}

\end{document}